\documentclass[10pt,a4paper]{article}
\title{Some BPS configurations of the BLG theory}
\author{
Shankhadeep Chakrabortty
\thanks{Institute of Physics, Bhubaneswar, India.
        Email:~\tt\email{sankha}{iopb.res.in}}
\and
Sudipto Paul Chowdhury
\thanks{Department of Theoretical Physics, Indian Association for the Cultivation of Science, Calcutta-700032, India.
        Email:~\tt\email{tpspc}{iacs.res.in}}
\and
Koushik Ray
\thanks{Department of Theoretical Physics, Indian Association for the Cultivation of Science, Calcutta-700032, India.
        Email:~\tt\email{koushik}{iacs.res.in}}
}
\newcommand\email[2]{{#1}@{#2}}
\usepackage{microtype,cmap,amsmath,amssymb,bm}
\usepackage{cite}
\usepackage{graphicx}
\usepackage[al]{pstricks}
\usepackage{pst-plot}
\usepackage[margin=3cm]{geometry}
\DeclareMathOperator{\tr}{Tr}

\newcommand{\ternary}[1]{{\langle\,#1\,\rangle}}

\newcommand{\R}{\mathbf{R}}
\newcommand{\G}{\Gamma}
\newcommand{\g}{\gamma}
\newcommand{\pa}{\partial}
\let\eq\eqref

\begin{document}

\maketitle

\begin{abstract}
\noindent We obtain BPS configurations of the BLG 
theory and its variant 
including mass terms for scalars and fermions in addition to
a background field with different world-volume and
$R$-symmetries. 
Three cases are considered,
with world-volume symmetries $SO(1,1)$ and $SO(2)$ and preserving different
amounts of supersymmetry.  In the former case 
we obtain a singular configuration preserving $N=(3,3)$ supersymmetry and 
an one-quarter BPS configuration corresponding to intersecting M2-M5-M5-branes.
In the latter instance the BPS equations are reduced to 
those in the self-dual Chern-Simons theory with two complex scalars.
In want of an exact solution, we find 
a topological vortex solution numerically in this case. 
Other solutions are given by combinations of domain walls.
\end{abstract}
\clearpage
\section{Introduction}
The Bagger-Lambert-Gustavsson(BLG) theory \cite{bagger1, bagger2,
bagger3, bagger4, gustav} is an ${N} = 8$, 
supersymmetric Chern-Simons type gauge theory 
based on a ternary gauge algebra coupled to matter in $(2+1)$-dimensions 
with $SO(8)$ $R$-symmetry. 
The theory is deemed to have $Osp(8|4)$ superconformal symmetry based on
strong evidences \cite{schwarz} 
and is thus a candidate for a world-volume theory of 
M2-branes. 
It contains eight scalar fields  interpreted as eight directions transverse 
to the world-volume of M2-branes in M-theory and 
eight corresponding fermions in addition to a gauge field. 
Imposing complete antisymmetry of the structure constant of the ternary 
gauge algebra  along with the closure of the supersymmetry
algebra constrains the gauge group to be $SO(4)$. 
The theory has a sixteen-dimensional moduli space lending itself to the
interpretation as a theory of two M2-branes.
\cite{lambert, mukhi1,ckpr}.
Various aspects and variants of the BLG theory have been considered 
\cite{mukhi2, mukhi3, mukhi4, raam, figu2, pass, cherkis, palmkvist, sudipto, chetan,lambert2, gustav2}.

In this article we shall be concened with BPS configurations in the BLG
theory and a particular modification of it. 
This entails the inclusion of mass terms for the 
scalars and the fermion and a flux term \cite{lambert2}.
BPS states of the BLG theory have been 
classified \cite{Kim} according to world-volume symmetries, namely,
$SO(1,2)$, $SO(1,1)$ and $SO(2)$.
A BPS configuration of the modified theory with $SO(1,2)$
world-volume symmetry and  $SO(4)$ $R$-symmetry has been studied earlier
\cite{ckpr}. 
In this article we study BPS configurations in 
three cases with $SO(1,1)$ and
$SO(2)$ world-volume symmetries. We consider BPS configurations in 
the BLG theory preserving $N=(3,3)$ supersymmetry 
with $SO(1,1)$ world-volume symmetry and 
$SO(3)\times SO(3)\times SO(2)$ $R$-symmetry and obtain a solution to the BPS
equations. The solution has scalars diverging at a finite distance of a
world-volume coordinate.
We then consider the
deformed variant with $SO(1,1)$ world-volume symmetry and
$SO(4) \times SO(4)$ $R$-symmetry preserving $N=(4,0)$ supersymmetry.
Finally we deal with an $N=4$ BPS configuration with 
$SO(2)$ world-volume symmetry and 
$SU(2) \times SO(4)$ $R$-symmetry. In the former case we obtain a
configuration which may be interpreted as a system of intersecting
M2-M5-M5-branes, following the popular interpretation of the BLG theory. The
other solutions turn out to be combination of domain walls interpolating
between pairs of classical vacua. In the latter
case the problem, upon choosing an appropriate ansatz, is mapped to the
self-dual $U(1)^2$ Chern-Simons theory with two complex scalar fields. This
problem has been studied earlier \cite{ziar,ccl,lpy}. In want of an analytic
solution to the BPS equations we present a numerical one corresponding to a
single vortex.  In considering these examples we
find that casting the BPS equations in terms of gauge-invariant 
variables used earlier \cite{ckpr} furnishes a useful guideline for the
choice of ansatze for scalars in the BPS equations. 

The article is organized as follows. 
In the following section we recall some aspects of the modified BLG theory.
In section~\ref{so3so3sol} we obtain solution to the BPS configurations with
$SO(1,1)$ symmetry in the world-volume and $SO(3)\times SO(3)\times SO(2)$
$R$-symmetry.
In section~\ref{domain:sol} we discuss the $SO(1,1)$-invariant BPS
configuration with $SO(4) \times SO(4)$ $R$-symmetry and 
its domain-wall solution. In section~\ref{so4:sol}  we proceed to discuss 
the $SO(2)$-invariant BPS configuration having $SU(2) \times SO(4)$ 
$R$-symmetry, map it to the Chern-Simons theory with two complex scalars 
and present a numerical solution for 
an Abelian topological vortex, before concluding in section~\ref{concl}. 
\section{BLG theory}\label{blg:rev}
Let us begin with a brief description of the BLG theory and
its deformation by a background four-form field. The modified
BLG theory is an ${N} = 8$ supersymmetric theory
in $2+1$-dimensions, given by the Lagrangian
\begin{equation}\label{action2}
  \mathcal{L} = \mathcal{L}_{\text{BLG}} + \mathcal{L}_\text{mass} +
  \mathcal{L}_\text{flux},
\end{equation} 
where the first term 
\begin{equation} 
\label{action1}
\begin{split}
\mathcal{L}_{\text{BLG}} &=  
-\frac{1}{2}\tr (D_{\mu} X^{I})(D^\mu X^{I}) 
 +\tr\frac{i}{2}\bar\Psi\gamma^\mu D_\mu \Psi
+  \frac{i}{4}\tr\bar\Psi \Gamma_{IJ} \ternary{X^I, X^J, \Psi} \\
&- \frac{1}{12}\tr\ternary{X^I, X^J, X^K}^2  
+  \frac{1}{2}\epsilon^{\mu\nu\lambda} 
\bigl( f^{abcd}A_{\mu ab}\partial_\nu A_{\lambda cd}
+ \frac{2}{3}f^{cda}{}_gf^{efgb} A_{\mu ab}A_{\nu cd}A_{\lambda ef} \bigr)
\end{split}
\end{equation} 
is the original BLG Lagrangian.
Here $\mu = 0, 1, 2$ designates the world-volume directions, $I =
1,\dotsc, 8$ indexes the flavors and $a = 1,2,3,4$ the gauge algebra.
$X^I_a$, $\Psi_a$ and $A_{\mu ab}$ are the scalars, the Majorana-Weyl
spinor and the gauge field, respectively. 
The three- and eight-dimensional gamma matrices are denoted
$\gamma$ and $\G$, respectively.  
The ternary bracket of the gauge algebra is denoted as $\langle~,~,~\rangle$,
while its structure constants are
denoted by $f^{abcd}$. Repeated indices are
summed over in the above expression and in the following unless stated
otherwise. Denoting the generators of the ternary algebra as
$\tau_a$, the metric tensor raising and lowering gauge indices is
written as
\begin{equation}
  h_{ab} = \tr \tau_a \tau_b.
\end{equation} 
We use the generators to write the fields valued in the ternary algebra as
\begin{gather}
\label{repscalar}
  X^I = h^{ab}X^I_a \tau_b, \\
\label{repfermion}
  \Psi = h^{ab}\Psi_a \tau _b.  
\end{gather}
Here $D_{\mu}$ denotes the covariant derivative, 
\begin{equation}
\label{covar}
D_{\mu} X^I_a = \partial_{\mu} X^I_a - \tilde{A_{\mu}}^b_a X^I_b
\end{equation}

In the presence of a four-form field $G_{IJKL}$
the BLG Lagrangian is augmented by a mass term
\begin{equation} 
\label{massterm}
\mathcal{L}_\text{mass} = -\frac{1}{2}m^2 \delta^{IJ} \tr(X^I X^J) 
+ c\tr(\overline{\Psi} \,  \Gamma^{I J K L} \, \Psi) 
\widetilde{G}_{IJKL}
\end{equation} 
and a flux term 
\begin{equation} 
\label{fluxterm}
\mathcal{L}_\text{flux} = - c \; \widetilde{G}_{I J K L} 
\tr(X^I \ternary{X^J, X^K, X^L}).
\end{equation} 
The four-form field
satisfies a self-duality condition
\begin{equation}\label{self-dual}
  \widetilde{G }_{IJKL}=G_{IJKL}, 
\end{equation} 
where the dual of the four-form field $G$ is defined as
\begin{equation}
 \widetilde{G}_{I J K L} = \frac{1}{4!} \epsilon_{I J K L P Q R S} \, G^{P Q R S}.
\end{equation}
The mass $m$ is determined by the four-form field as
$m^2 = \frac{c^2}{768} G^2$, with $G^2 = G^{IJKL} G_{IJKL}$ and 
$c$ is a parameter which is found to be equal to 2 \cite{lambert2}.
Thus the BLG theory is recovered in the limit of vanishing $c$.

The action corresponding to the Lagrangian \eqref{action1} is 
under the supersymmetry transformations \cite{lambert2}
\begin{gather}
\label{susyscalar}
  \delta X^I = i \; \overline{\theta} \; \Gamma^I \Psi, \\
\label{susyfermion}
  \delta \Psi = \gamma^{\mu} \Gamma^I D_{\mu} X^I \theta - 
  \frac{1}{6} \Gamma^{I J K} \ternary{X^I, X^J, X^K} \, \theta + 
  \frac{c}{8} \Gamma^{I J K L} \Gamma^M \widetilde{G}_{I J K L} X^M \theta\\
\label{susygauge}
  \delta A_{\mu}(\phi) = i \; \overline{\theta} \, \gamma_\mu \Gamma^I \ternary{\Psi, X^I, \phi},
\end{gather}
where $\phi$ in the transformation of the gauge field
represents either a $X^I$ or $\Psi$ and $\theta$ denotes the
parameter of supersymmetry variation,
satisfying
\begin{equation}
\label{gammaproj}
\begin{gathered}
\Gamma^9=\Gamma^{1\ldots 8}\;\theta=\theta\\
\gamma^{txy}\theta=\theta. 
\end{gathered}
\end{equation} 
The supersymmetry
transformations close on-shell up to translation and local gauge
transformations if  the structure constant of the ternary algebra is
the rank-four antisymmetric tensor, that is, 
\begin{equation}\label{strucconst}
f^{a b c d} = \epsilon^{a b c d},
\end{equation} 
so that the gauge group is $SO(4)$.
The scalars and the fermion transform as vectors of the gauge group $SO(4)$.
We thus choose, for example, 
\begin{equation}
X^I = \begin{pmatrix}
X^I_1\\X^I_2\\X^I_3\\X^I_4
\end{pmatrix},
\end{equation}
for all $I$. 
For future convenience we have set the level of the Chern-Simons 
action to be unity, and the metric is taken to be Euclidean, 
\begin{equation}\label{symbilin}
h_{ab} = \delta_{ab}.
\end{equation}
The ternary bracket then reads
\begin{equation}\label{algebra}
\ternary{X^I,X^J,X^K} = \epsilon^{abcd} X^I_a X^J_b X^K_c \tau_d.
\end{equation}
We shall be concerned with the BPS configurations of the theory with
Lagrangian \eqref{action2}.
The BPS equation is obtained by setting the supersymmetry variation of the
fermion to zero, that is
\begin{equation}
\delta\Psi = 0.
\end{equation}
Depending on the subgroup of the $R$-symmetry as well as the
world-volume symmetry to be maintained, the supersymmetry parameter $\theta$
is restricted by means of a projector, $\Omega$. Thus, the BPS equations
are given by 
\begin{equation}\label{BPSeqn}
\bigl[D_\mu X^I\gamma^\mu \Gamma^I 
-\frac{1}{6}\G^{IJK}\ternary{X^I, X^J, X^K}
+\frac{c}{8}\Gamma^{IJKL}\G^M \widetilde{G}_{I J K L}X^M \bigr]
\Omega \theta = 0.
\end{equation}
Let us note that only the anti-self-dual combination of the four-form
field appears in the last term on the left hand side, linear in $X$. 
The $R$-symmetry in this formulation is realized explicitly in
terms of the four-form field as
\begin{equation}\label{Rsymm}
R^I_J = \overline{\theta_2} \; \Gamma^{IKLM} \, \theta_1 \, 
\widetilde{G}_{KLMJ},
\end{equation}
where $\theta_1$ and $\theta_2$ are two parameters of supersymmetry
variation.
The conserved charged under the global $SO(8)$ symmetry of the BLG theory is given by the $R$-charge, namely
\begin{equation}
\label{rcharge}
R^{IJ} = \int d^2x \left(X^{Ia} D_0 X^J_a - X^{Ja} D_0 X^I_a + \frac{i}{2}\overline{\psi}^a\g^0 \G^{IJ}\psi_a\right)
\end{equation}
where $R^{IJ}$ is antisymmetric in $I$ and $J$.
We now proceed to study certain BPS configurations of the theory discussed
above. 

Furthermore, the BPS configurations have to satisfy the Gauss constraint,
namely
\begin{equation}
\label{gaussfull}
{F_{\mu\nu}} ^a_b + \epsilon_{\mu\nu\lambda} \epsilon^{cda}_{~~b}
X^I_cD^{\lambda}X^I_d =0,
\end{equation} 
where $F$ denotes the field strength corresponding to the gauge field
$\tilde{A}$. 
\section{BPS configuration with $SO(1,1)\times SO(3)\times SO(3)
\times SO(2)$ symmetry}
\label{so3so3sol}
In this section we present a solution to the BPS equations
preserving $N=(3,3)$ supersymmetry in the BLG theory without the mass and the
four-form  terms, corresponding to the Lagrangian \eq{action1}. 
The world-volume has $SO(1,1)$ symmetry and the
$R$-symmetry is $ SO(3)\times SO(3)\times SO(2)$. The equations are \cite{Kim}
\begin{equation} 
\label{so3DX}
D_t X^I =0, \quad D_xX^I=0, 
\end{equation} 
with
\begin{equation}
\begin{gathered}
\label{so3bps}
\mathcal{J}_p^{IJ}D_yX_J
+\frac{1}{2}\mathcal{J}_p^{JK} \ternary{X^I,X_J,X_K}=0,\\
{\mathcal{J}}_{p+3}^{IJ}D_yX_J
-\frac{1}{2}{\mathcal{J}}_{p+3}^{JK}\ternary{X^I,X_J,X_K}=0,
\end{gathered}
\end{equation} 
where $p=1,2,3$, $I,J=1,2,\cdots , 8$ and 
the complex structures $\mathcal{J}$ are defined by 
\begin{equation} 
\begin{gathered}
\frac{1}{2}\mathcal{J}_1^{IJ} \Gamma_{IJ} =
\Gamma^{12}+\Gamma^{34}+\Gamma^{56}+\Gamma^{78}, \quad
\frac{1}{2}\mathcal{J}_4^{IJ} \Gamma_{IJ} =
\Gamma^{12}+\Gamma^{43}+\Gamma^{56}+\Gamma^{87},\\
\frac{1}{2}\mathcal{J}_2^{IJ} \Gamma_{IJ} =
\Gamma^{14}+\Gamma^{23}+\Gamma^{58}+\Gamma^{67}, \quad
\frac{1}{2}\mathcal{J}_5^{IJ} \Gamma_{IJ} =
\Gamma^{17}+\Gamma^{28}+\Gamma^{53}+\Gamma^{64},\\
\frac{1}{2}\mathcal{J}_3^{IJ} \Gamma_{IJ} =
\Gamma^{13}+\Gamma^{42}+\Gamma^{57}+\Gamma^{86},\quad
\frac{1}{2}\mathcal{J}_6^{IJ} \Gamma_{IJ} =
\Gamma^{18}+\Gamma^{72}+\Gamma^{54}+\Gamma^{36}.
\end{gathered}
\end{equation} 
Thus from \eq{so3bps} we have six expressions for each $D_yX^I$, for
$I=1,2,\cdots , 8$. Comparing the various expressions for the same $D_yX^I$
we obtain a set of relations among the ternary brackets, namely,
\begin{equation}
\label{reln1}
\begin{gathered}
\scriptstyle
\ternary{X^2,X^5,X^6}=0,\quad
\ternary{X^3,X^5,X^7}=\ternary{X^4,X^5,X^8},\quad
\ternary{X^3,X^6,X^8}=-\ternary{X^4,X^6,X^7}, \\
\scriptstyle
\ternary{X^1,X^5,X^6}=0,\quad
\ternary{X^4,X^6,X^8}=\ternary{X^3,X^6,X^7},\quad
\ternary{X^4,X^5,X^7}=-\ternary{X^3,X^5,X^8}, \\
\scriptstyle
\ternary{X^4,X^7,X^8}=0,\quad
\ternary{X^2,X^6,X^7}=\ternary{X^1,X^5,X^7},\quad
\ternary{X^2,X^5,X^8}=-\ternary{X^1,X^6,X^8}, \\
\scriptstyle
\ternary{X^3,X^7,X^8}=0,\quad
\ternary{X^2,X^6,X^8}=\ternary{X^1,X^5,X^8},\quad
\ternary{X^2,X^5,X^7}=-\ternary{X^1,X^6,X^7}, \\
\scriptstyle
\ternary{X^1,X^2,X^6}=0,\quad
\ternary{X^1,X^4,X^8}=\ternary{X^1,X^3,X^7},\quad
\ternary{X^2,X^4,X^7}=-\ternary{X^2,X^3,X^8}, \\
\scriptstyle
\ternary{X^1,X^2,X^6}=0,\quad
\ternary{X^2,X^4,X^8}=\ternary{X^2,X^3,X^7},\quad
\ternary{X^1,X^4,X^7}=-\ternary{X^1,X^3,X^8}, \\
\scriptstyle
\ternary{X^3,X^4,X^8}=0,\quad
\ternary{X^2,X^3,X^6}=\ternary{X^1,X^3,X^5}, \quad
\ternary{X^2,X^4,X^5}=-\ternary{X^1,X^4,X^6},\\
\scriptstyle
\ternary{X^3,X^4,X^7}=0,\quad
\ternary{X^2,X^4,X^6}=\ternary{X^1,X^4,X^5}, \quad
\ternary{X^2,X^3,X^5}=-\ternary{X^1,X^3,X^6},\\
\end{gathered}
\end{equation} 
\begin{equation} 
\label{reln2}
\begin{gathered}
\scriptstyle
\ternary{X^2,X^3,X^4}=\ternary{X^2,X^7,X^8}
=\ternary{X^4,X^5,X^8}+\ternary{X^4,X^6,X^7},\quad
\ternary{X^1,X^3,X^4}=\ternary{X^1,X^7,X^8}
=-\ternary{X^3,X^5,X^8}-\ternary{X^3,X^6,X^7},\\
\scriptstyle
\ternary{X^1,X^2,X^4}=\ternary{X^4,X^5,X^6}
=\ternary{X^1,X^6,X^8}-\ternary{X^1,X^5,X^7},\quad
\ternary{X^1,X^2,X^3}=\ternary{X^3,X^5,X^6}
=\ternary{X^1,X^5,X^8}+\ternary{X^1,X^6,X^7},\\
\scriptstyle
\ternary{X^3,X^4,X^6}=\ternary{X^6,X^7,X^8}
=\ternary{X^1,X^3,X^7}+\ternary{X^2,X^3,X^8},\quad
\ternary{X^5,X^7,X^8}=\ternary{X^3,X^4,X^4}
=\ternary{X^1,X^3,X^8}-\ternary{X^2,X^3,X^7},\\
\scriptstyle
\ternary{X^1,X^2,X^8}=\ternary{X^5,X^6,X^8}
=-\ternary{X^1,X^4,X^6}-\ternary{X^1,X^3,X^5},\quad
\ternary{X^1,X^2,X^7}=\ternary{X^5,X^6,X^7}
=\ternary{X^1,X^4,X^5}-\ternary{X^1,X^3,X^6},
\end{gathered}
\end{equation} 
together with a set of Basu-Harvey equations with respect to the world-volume
coordinate $y$,
\begin{equation}
\begin{gathered}
\label{bhset1}
D_yX^1 =  2\ternary{X^2,X^3,X^4},\quad
D_yX^2 = -2\ternary{X^1,X^3,X^4}, \\
D_yX^3 =  2\ternary{X^1,X^2,X^4},\quad
D_yX^4 = -2\ternary{X^1,X^2,X^3}, \\
D_yX^5 =  2\ternary{X^3,X^4,X^6},\quad
D_yX^6 = -2\ternary{X^5,X^7,X^8}, \\
D_yX^7 =  2\ternary{X^1,X^2,X^8},\quad
D_yX^8 = -2\ternary{X^1,X^2,X^7}. \\
\end{gathered}
\end{equation} 
It will be useful to first write the BPS equations in terms of gauge-invariant
variables \cite{ckpr}. This furnishes a guideline for the choice of 
ansatze for the $X$'s. Let us introduce the gauge-invariant fields
\begin{equation}
Y^{IJ}=\sum_{a =1}^4 X^{Ia}X^J_a,
\end{equation}  
where indices are raised or lowered with
the Euclidean bilinear \eqref{symbilin}. The gauge-invariants satisfy
\begin{equation}
\label{delY}
\partial_{\mu} Y^{IJ} = X^{Ia}D_{\mu}X^J_a + X^{Ja}D_{\mu}X^I_a
\end{equation} 
due to the antisymmetry of the gauge field. 
Using this and \eq{reln1} and \eq{reln2} we obtain from \eq{bhset1} 
a set of first-order
equations for the gauge-invariants
\begin{equation} 
\begin{gathered}
\label{bhgi}
\pa_yY^{11} = - 4 F_{1234}, \quad
\pa_yY^{15} = -2 F_{4567}-2F_{1238}, \quad
\pa_yY^{16} = -2 F_{4568}+2F_{1237}, \\
\pa_yY^{22} = - 4 F_{1234}, \quad
\pa_yY^{25} = -2 F_{1345}-2F_{2346}, \quad
\pa_yY^{26} = -2 F_{1346}+2F_{2345}, \\
\pa_yY^{33} = - 4 F_{1234}, \quad
\pa_yY^{37} = 2 F_{1247}+2F_{1238}, \quad
\pa_yY^{38} = 2 F_{4568}-2F_{5678}, \\
\pa_yY^{44} = - 4 F_{1234},\quad
\pa_yY^{47} = 2 F_{1248}, \quad
\pa_yY^{48} = 2 F_{1247}\\
\pa_yY^{55} = - 4 F_{3456},\quad
\pa_yY^{66} = - 4 F_{3456},  \\
\pa_yY^{77} = - 4 F_{1278},\quad
\pa_yY^{88} = - 4 F_{1278}, 
\end{gathered}
\end{equation} 
and $\pa_yY^{IJ}=0$ for all other $I$, $J$.
So far our analysis has been completely general. Now Let us assume that the
constant $Y$'s, not appearing in \eq{bhgi}, are zero. Also, from the above
set of first-order equations we note that it is convenient to choose 
\begin{equation}
X^i = f S^i, 
\end{equation}
where $i=1,2,3,4$ and $S^i$ are the four mutually orthogonal 
canonical basis vectors of $\R^4$.
Then from \eq{bhgi} it is apparent that 
$X^5$ and $X^6$ are linear combinations of $S^1$ and $S^2$,
while $X^7$ and $X^8$ are linear combinations of $S^3$ and $S^4$. Using
the relations \eq{reln1} and \eq{reln2} we can fix the coefficients of these
linear combinations and obtain 
\begin{equation}
\begin{gathered}
\label{presoln}
X^i =f S^i,\quad i=1,2,3,4,\\ 
X^5 = f S^1 \cos\theta + fS^2 \sin\theta,\quad 
X^6 = - fS^1 \sin\theta + fS^2 \cos\theta, \\
X^7 = fS^3 \cos\theta + fS^4 \sin\theta, \quad
X^8 = -fS^3 \sin\theta + fS^4 \cos\theta,
\end{gathered}
\end{equation}  
where $f$ and $\theta$ are to be determined from the differential equations. 

From the equation for $Y^{11}$, say, we obtain an equation for $f$,
\begin{equation}
\pa_y f = -2f^3, 
\end{equation} 
which is solved to obtain 
\begin{equation}
f^2 = \frac{1}{c+4y}, 
\end{equation} 
where $c=2$ and we have chosen the integration constant to be vanishing. 
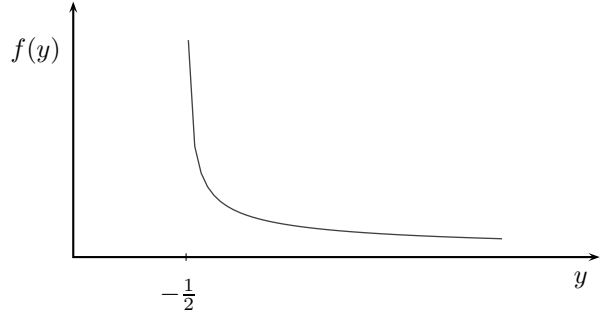
\begin{figure}[h]
\begin{center}
\begin{pspicture}(-1,0)(5,3.6)
\psset{unit=10mm}
\psline{<->}(-2,3.4)(-2,0)(5,0)
\psplot[linecolor=darkgray,linewidth=.5pt]{-.47}{3.7}{1 2 4 x mul add sqrt div}
\rput(4.75,-.3){$y$}
\rput(-2.5,2.75){$f(y)$}
\rput(-.6,-.5){$-\frac{1}{2}$}
\psline[linewidth=.4pt](-.5,.05)(-.5,-.05) 
\end{pspicture}
\end{center}
\caption{Plot of $f(y)$ for the $N=(3,3)$ configuration}
\label{33:fig}
\end{figure}
Now, comparing the expressions for $D_y X^5$, $D_y X^1$ and $D_yX^2$, upon 
using \eq{presoln}, we obtain 
\begin{equation}
\pa_y \log\cos\theta = 0, 
\end{equation} 
implying that $\theta$ is a constant. 
We have thus fixed the solution \eq{presoln} completely. 
A sketch of the function $f(y)$ is shown in Figure~\ref{33:fig}.
For this solution all the scalars diverge at a finite distance in the $y$
direction, namely, 
$y=-1/2$. Interpretation of this solution in terms of M2- and M5-branes
is not clear. 
\section{BPS configurations with $SO(1,1)\times SO(4)\times SO(4)$ symmetry}
\label{domain:sol}
In this section we study $SO(1,1)$-invariant $N=(4,0)$ BPS configurations 
having $SO(4)\times SO(4)$ $R$-symmetry. The BPS
equations are derived for the modified BLG theory by applying the 
appropriate BPS projection operator on the supersymmetry variation of the
fermion and equating it to zero by \eqref{BPSeqn}.
The generic form of the $SO(1,1)$ BPS projector is given 
in terms of $32 \times 32$ gamma matrices as  
\begin{equation}
\label{project}
\begin{split}
\Omega = &\frac{1}{16}(1+ \alpha_0 \g^{t x})(1 - \alpha_1 \alpha_2 \G^{1278} + \alpha_1 \alpha_3 \G^{1368} - 
\alpha_1 \G^{2468} - \alpha_3 \G^{3478}\\ 
&- \alpha_2 \G^{5678} + \alpha_1 \alpha_2 \alpha_3 \G^{2358} + \alpha_2 \alpha_3 \G^{1458})\mathcal{P} 
\end{split}
\end{equation}
where $\alpha_0, \alpha_1, \alpha_2, \alpha_3$ are sign factors
assuming values $\pm 1$ and $\g$ and $\G$  designates the gamma matrices 
defined on the world-volume and transverse to the world-volume, respectively.
The chiral projection operator $\mathcal{P}$ is 
defined in terms of $\G^9 = \G^{12\cdots 8}$ as
\begin{equation}
\label{chiproject}
\mathcal{P} = \frac{1}{2}(1 + \G^9). 
\end{equation}
Different choices of the sign factors in (\ref{project}) 
give the BPS projection matrix which correspond to 
breaking $R$-symmetry in a certain manner \cite{Kim, Bak, loginov}. 
The projector preserving $SO(4)\times SO(4)$ $R$-symmetry is obtained by
summing the four $N=1$ projectors \eq{project} with 
the choice of $\alpha$'s as
$\{+,+,+,+\}$, $\{+,+,+,-\}$, $\{+,+,-,+\}$ , 
$\{+,+,-,-\}$ and  is given by      
\begin{equation}
\label{bpsso4}
\Omega = \frac{1}{4}(1+ \g^{t x})(1 + \G^{5678}) (1+ \G^9).
\end{equation}
By equation \eq{gammaproj} this operator corresponds to the projections
\begin{equation}
\label{proj1}
\begin{gathered}
\g^{txy}\theta=\theta, \quad
\g^{tx}\Gamma^{1234}\theta = \theta,\quad
\g^{tx}\Gamma^{5678}\theta = \theta,
\end{gathered} 
\end{equation} 
the last one being a dependent one.
Applying the projection matrix (\ref{bpsso4}) on (\ref{BPSeqn}), we obtain
the BPS equations. These comprise of differential equations namely
\begin{equation}
\label{bpsso41}
(D_{t} - D_{x}) X^I = 0,
\end{equation}
for all $I=1,2,\cdots , 8$ and a set of modified Basu-Harvey equations,
\begin{gather} 
\label{SO41}
D_{y} X^i - \frac{1}{6}\epsilon^{ijkl}\ternary{X_j,X_k,X_l}  
-\eta_1 X^{i} = 0,\\
\label{SO42}
D_{y} X^p - \frac{1}{6}\epsilon^{pqrs}\ternary{X_q,X_r,X_s} 
- \eta_2 X^{p} = 0,
\end{gather} 
where we have split the flavor indices as $i,j,k,l=1,2,3,4$ and 
$p,q,r,s=5,6,7,8$. $D_t$, $D_x$ and $D_y$ designate  
the covariant derivatives with respect to the world-volume
coordinates (\ref{covar}). 
Coefficients of the terms linear in 
$X$'s in equation \eqref{SO41}
are determined in terms of the four-form field by $\eta_1=3 c G_{1234}$ 
and $\eta_2= 3 c G_{5678}$ with $c = 2$, as discussed  above. 
We assume $\eta_1$ and $\eta_2$ to be positive in the following.


Our objective here is to find a topological
solution to the BPS equations \eqref{bpsso41}, \eqref{SO41} and
\eqref{gaussfull}. First let us write down the BPS equations in terms of the 
gauge-invariant variables $Y$  as before. 
Multiplying both sides of \eq{SO41} and \eq{SO42}
with an appropriate $X$ and taking linear combinations, 
using \eq{delY}, we obtain equations for the gauge-invariants
\begin{equation} 
\begin{gathered} 
\label{Yeq1}
\partial_y Y^{ij}-2\eta_1Y^{ij}=-2\delta^{ij} F_{1234} ,\\
\partial_y Y^{rs}-2\eta_2Y^{rs}=-2\delta^{rs} F_{5678} ,\\
\partial_y Y^{ip} - \frac{1}{3!}(\varepsilon^{ijkl}F_{jklp} - \varepsilon^{pqrs}F_{iqrs}) 
- (\eta_1+\eta_2)Y^{ip} = 0
\end{gathered}
\end{equation}  
for $i,j=1,\cdots ,4$ and $r,s=5,6,7,8$, 
where we defined the gauge-invariant four-form
$F_{IJKL} = X^I_a X^J_b X^K_c X^L_d \epsilon^{abcd}$.

Further, using \eqref{bpsso41} in the Gauss constraint
\eqref{gaussfull} we get
\begin{equation} 
F_{xy}=F_{ty},
\end{equation}
while using \eqref{SO41} and \eq{SO42} to eliminate the covariant
$y$-derivatives in \eqref{gaussfull} we obtain
\begin{equation}
\label{ftx}
F_{tx}=0. 
\end{equation}  
The modified Basu-Harvey equations 
have been solved earlier to obtain a
domain wall and a fuzzy funnel solution \cite{bagger3,chu}. Let us
point out that the
equations for the scalars are the same for half and quarter BPS configurations 
with $SO(4)\times SO(4)$ 
$R$-symmetry. The gauge fields satisfy different equations in
these two cases, however. For example, while we have the equation
\eq{bpsso41} for the quarter-BPS configuration, the half-BPS configurations
satisfy $D_tX^I=D_xX^I=0$ \cite{Kim}. 
Hence, it is important to write down the gauge
fields explicitly, even if in a special gauge. 
In order to
obtain explicit expressions for gauge field configurations
we shall choose simplifying ansatze. 
The scalar fields $X^i$, $i=1,2,3,4$, are chosen to be mutually
orthogonal $SO(4)$ vectors, as are $X^p$, $p=5,6,7,8$. However, if we
choose the basis vectors to be the constant vectors $S^i$ as in the last
section, the gauge fields will remain undetermined. Hence for the present
case we choose a different set of mutually orthogonal basis vectors and
express the scalars as
\begin{equation} 
\begin{gathered}
\label{scalaransatz}
X^1 = f\begin{pmatrix}\cos\Theta\\
\sin\Theta\\0\\0\end{pmatrix},\;
X^2 = \kappa f\begin{pmatrix}- \sin\Theta\\
\cos\Theta\\0\\0\end{pmatrix},\;
X^3 =  f\begin{pmatrix}0\\0\\\cos\Phi\\ 
\sin\Phi\end{pmatrix},\;
X^4 =  f\begin{pmatrix}0\\0\\- \sin\Phi\\ 
\cos\Phi\end{pmatrix},\\
X^5 = g\begin{pmatrix}\cos\Theta\\
\sin\Theta\\0\\0\end{pmatrix},\;
X^6 = \kappa' g\begin{pmatrix}- \sin\Theta\\
\cos\Theta\\0\\0\end{pmatrix},\;
X^7 =  g\begin{pmatrix}0\\0\\\cos\Phi\\ 
\sin\Phi\end{pmatrix},\;
X^8 =  g\begin{pmatrix}0\\0\\- \sin\Phi\\ 
\cos\Phi\end{pmatrix},
\end{gathered}
\end{equation} 
where $\kappa,\kappa' = \pm1$. We shall find solutions corresponding 
to both signs of $\kappa$, $\kappa'$. The gauge-invariant co-ordinates for this choice are
$Y^{ii}=f^2$, $i=1,2,3,4$, 
$Y^{pp}=g^2$, $p=5,6,7,8$. 
The gauge field is chosen to be of the form 
\begin{equation}
\label{gfield}
\tilde{A}_{\mu} = \left(\begin{array}{cccc} 0 & 
\tilde{A_{\mu}}^1_2 & 0 & 0 \\ 
- \tilde{A_{\mu}}^1_2 & 0 & 0 & 0 \\ 0 & 0 & 0 & \tilde{A_{\mu}}^3_4 
\\ 0 & 0 & - \tilde{A_{\mu}}^3_4 & 0\end{array}\right),
\end{equation}
in accordance with the aove choice for the scalars,
hence breaking the gauge group $SO(4)$ to $SO(2) \times SO(2)$.
In equation \eqref{scalaransatz} the functions $\Theta$ and $\Phi$ 
correspond to the freedom of gauge choice for the
residual $SO(2)\times SO(2)$ subgroup.   
For future convenience we shall leave these arbitrary. 
However, the stress-energy tensor does not depend on them, as we find later. 

We shall first obtain restrictions on $f,g,\Theta,\Phi$ and then determine
the components of the gauge field in terms of them.
We restrict ourselves to stationary solutions. Then all the fields are
independent of time. 
From \eqref{Yeq1} we obtain equations for $f$ and $g$, namely
\begin{equation}
\begin{gathered}
\label{feqn}
\partial_yf =\eta_1f -\kappa f^3,\\
\partial_yg =\eta_2g -\kappa' g^3.
\end{gathered}
\end{equation} 
We now relate the components of the gauge field to $f,g,\Theta,\Phi$.
First, using \eqref{scalaransatz} and the equation for $X^1$ from 
\eqref{SO41} we obtain, for the first component,
\begin{equation}
(\partial_yf-\eta_1f +\kappa f^3)\cos\Theta 
- f (\tilde{A_y}^1_2-\pa_y\Theta)\sin\Theta =0,
\end{equation} 
and similarly for the first component of the  equation for $X^3$ from 
\eq{SO41}. 
Comparing with \eq{feqn} we obtain 
\begin{equation} 
\label{Ayeq}
{\tilde{A_y}}^1_2 = \partial_y\Theta,\qquad
{\tilde{A_y}}^3_4 = \partial_y\Phi.
\end{equation} 
This demonstrates the utility of the gauge-invariant equations and justifies
keeping $\Theta$ and $\Phi$ arbitrary in \eq{scalaransatz}. 
Setting the temporal derivative to zero in \eqref{bpsso41}  
we obtain 
\begin{equation}
\label{gauge1}
(\tilde{A_t}^b_a  - \tilde{A_x}^b_a) X^I_b = - \partial_{x} X^I_a
\end{equation}
Putting $a=1,2$, resp. $b=2,1$ and using 
$\tilde{A_{\mu}}^a_b = - \tilde{A_{\mu}}^b_a$ and equations
\eqref{scalaransatz}, we obtain
\begin{equation}
\partial_x f=0,
\end{equation} 
that is, $f$ is independent of $x$. Similarly, from \eq{SO42} and \eq{feqn}
we obtain $\pa_x g=0$. In other words, $f$ and $g$ are functions of $y$
only. This leads to 
\begin{equation}
\label{gagcond1}
\begin{split}
\tilde{A_t}^1_2  - \tilde{A_x}^1_2 &= - \partial_x \Theta \\
\tilde{A_t}^3_4  - \tilde{A_x}^3_4 &= - \partial_x \Phi. 
\end{split}
\end{equation}
Thus the gauge fields are determined in terms of  
$\Theta$ and $\Phi$. 
Using \eqref{Ayeq}, \eq{bpsso41} and \eqref{gagcond1}, the Gauss constraint
\eqref{gaussfull} yields
\begin{equation} 
\begin{gathered}
\label{gauss4}
\partial_y {\tilde{A_t}}^1_2 = 2 (f(y)^2+g(y)^2){\tilde{A_t}}^3_4,\\
\partial_y {\tilde{A_t}}^3_4 = 2 (f(y)^2+g(y)^2){\tilde{A_t}}^1_2,
\end{gathered}
\end{equation} 
while \eqref{ftx} yields 
\begin{equation} 
\label{gauss6}
\partial_x {\tilde{A_t}}^1_2 = 0 = \partial_x {\tilde{A_t}}^3_4 ,
\end{equation} 
so that $\tilde{A}_t$ is a function of $y$ alone. 
Now, eliminating the combination
$f^2+g^2$ between the two equations in \eqref{gauss4} we obtain 
\begin{equation}
\label{sqcomb}
\partial_y \big(({\tilde{A_t}}^1_2)^2 - ({\tilde{A_t}}^3_4)^2 \big)=0, 
\end{equation} 
leading to the conclusion 
that the squares of the components of the gauge field $A_t$ may
differ by a constant only. 
By linearly combining the equations \eqref{gauss4} we can
cast them as first order differential equations for the combinations
${\tilde{A_t}}^1_2 \pm {\tilde{A_t}}^3_4$ as
\begin{equation} 
\partial_y({\tilde{A_t}}^1_2 \pm {\tilde{A_t}}^3_4) = 
\pm 2(f^2+g^2)({\tilde{A_t}}^1_2 \pm {\tilde{A_t}}^3_4), 
\end{equation} 
which are solved to obtain
\begin{equation} 
\begin{gathered}
\label{Atsoln}
{\tilde{A_t}}^1_2 = A_0 e^{\int (f^2+g^2)dy }+B_0e^{-\int (f^2+g^2)dy},\\
{\tilde{A_t}}^3_4 = A_0 e^{\int (f^2+g^2)dy }-B_0e^{-\int (f^2+g^2)dy},
\end{gathered}
\end{equation}  
where $A_0$ and $B_0$ are constants. This solution satisfies \eqref{sqcomb}.
The solution to  \eqref{feqn} is
\begin{equation}
\label{modf}
f(y) = \pm \frac{\sqrt{\eta_1} a}{\sqrt{e^{- 2 \eta_1 y} + \kappa a^2}},\quad 
g(y) = \pm \frac{\sqrt{\eta_2} a'}{\sqrt{e^{- 2 \eta_2 y} + \kappa' a'^2}}, 
\end{equation}
where $a,a'$ are  constants of integration. 

Different configurations ensue from the choices of $\kappa, \kappa'$. 
By allowing the supersymmetry variation of $\mathcal{L}$ to
vanish \cite{lambert2}, the mass parameter in the scalar term in 
$\mathcal{L}_{\text{mass}}$ gets related to the
four-form field $\tilde{G}_{IJKL}$ as
\begin{equation}
\label{mass}
\G^{IJKL}\tilde{G}_{IJKL}\G^{MNOP}\tilde{G}_{MNOP} = \frac{32
m^2}{c^2}\big(1 + \G^{12345678}\big),
\end{equation}
resulting in $m^2= 9 c^2 G^2$, where 
$G^2 = G_{1234}G^{1234} = G_{5678}G^{5678}$.   
Using this value of $m$  and the ansatz for the scalars \eqref{scalaransatz},
the scalar potential obtained from the sextic term in
$\mathcal{L}_{\text{BLG}}$, the scalar term in $\mathcal{L}_{\text{mass}}$
along with $\mathcal{L}_{\text{flux}}$ is
\begin{equation}
\label{potsol}
\begin{split}
V &= - \frac{1}{12}\tr\ternary{X^I, X^J, X^K}^2  
-\frac{1}{2}m^2 \delta^{IJ} \tr(X^I X^J) 
- c \; \widetilde{G}_{I J K L} 
\tr(X^I \ternary{X^J, X^K, X^L})\\
&= -2f^2(f^2 - {\eta_1})^2 -2g^2(g^2 - \eta_2)^2, 
\end{split}
\end{equation}
with $\eta_1 , \eta_2 > 0$. 
The classical vacua are therefore,
\begin{equation}
\begin{gathered}
\label{vacua}
V_{I}:~f(y)=g(y) = 0;\quad 
V_{II}:~f(y) = \pm \sqrt{\eta_1},\; g(y) = \pm \sqrt{\eta_2}\\
V_{III}:~f(y)=0,\; g(y) = \pm \sqrt{\eta_2},\quad
V_{IV}:~f(y) = \pm \sqrt{\eta_1},\; g(y)=0.
\end{gathered} 
\end{equation} 

The solution \eqref{modf} with $\kappa=\kappa'=1$
interpolates between the vacua $V_I$ and
$V_{II}$ as $y$ varies from $-\infty$ to $\infty$. Taking into
account that the solution is independent of $x$, this is therefore a domain
wall solution. 
By \eqref{Atsoln}, the temporal component of the gauge field is 
\begin{equation} 
\begin{gathered}
\label{solgauge1}
{\tilde{A_t}}^1_2 = A_0 \sqrt{(1 + a^2 e^{2 \eta_1 y})(1 + a'^2 e^{2 \eta_2 y})} + 
\frac{B_0}{\sqrt{(1 + a^2 e^{2 \eta_1 y})(1 + a'^2 e^{2 \eta_2 y})}},\\
{\tilde{A_t}}^3_4 =A_0 \sqrt{(1 + a^2 e^{2 \eta_1 y})(1 + a'^2 e^{2 \eta_2 y})} - 
\frac{B_0}{\sqrt{(1 + a^2 e^{2 \eta_1 y})(1 + a'^2 e^{2 \eta_2 y})}}          
\end{gathered}
\end{equation} 
However, in order to keep $\tilde{A_t}$ finite in the whole domain of 
$y$, we have to set $A_0$ to zero. Thus, finally, the two components of 
$\tilde{A_t}$ are given by 
\begin{equation}
\label{gauge:fin}
{\tilde{A_t}}^1_2 
= -{\tilde{A_t}}^3_4
= \frac{B_0}{\sqrt{(1 + a^2 e^{2 \eta_1 y})(1 + a'^2 e^{2 \eta_2 y})}},
\end{equation}
where we have taken the difference of their squares to be vanishing. 
Having thus obtained $\tilde{A_t}$, $\tilde{A_x}$ and $\tilde{A_y}$ 
are determined, up to gauge
transformation, by \eqref{gagcond1} and \eqref{Ayeq}, respectively. 
The enrgy-momemtum tensor is obtained by varying the Lagrangian
$\mathcal{L}$ with respect to the world-volume metric,
\begin{equation}
\label{enmom}
T_{\mu \nu} = \frac{1}{2} D_{\mu}X^{Ia}D_{\nu}X^I_a 
- \frac{1}{4} g_{\mu\nu}\big(D^{\alpha}X^{Ia}D_{\alpha}X^I_a - V \big),
\end{equation}
where the potential $V$ is given by \eq{potsol}.
Energy density of the configuration obtained above is given by
$T_{tt}$.
Plugging in the solutions for the scalars, \eqref{modf}, 
and the gauge fields, \eqref{Ayeq}, 
\eqref{gagcond1} and \eqref{gauge:fin} in the expression for 
$T_{tt}$, we obtain the energy density to be   
\begin{equation}
\label{energy}
T_{tt} = 2\eta_1 e^{2 \eta_1 y}
\left(\frac{1+ 3 \eta_1^2 ( 4- a^2 e^{2 \eta_1 y} -1)^2}{( 1 + a^2 e^{2
\eta_1 y})^3}\right) + 2\eta_2 e^{2 \eta_2 y}
\left(\frac{1+ 3 \eta_2^2 ( 4 - a'^2 e^{2 \eta_2 y} -1)^2}{( 1 + a'^2 e^{2
\eta_2 y})^3}\right).
\end{equation}

For $\kappa=\kappa'=-1$, the $f$ and $g$ and hence the gauge invariants
$Y^{ii}$ and $Y^{rr}$ diverge at 
$y= -\frac{1}{\eta_1}\ln{a}$  and 
$y= -\frac{1}{\eta_2}\ln{a'}$, respectively.
The corresponding solutions for the gauage fields ${\tilde{A_t}}^1_2$ and ${\tilde{A_t}}^3_4$
are given by 
\begin{equation} 
\begin{gathered}
\label{gaugesol2}
{\tilde{A_t}}^1_2 = C_0 \sqrt{(1 - a^2 e^{2 \eta_1 y})(1 - a'^2 e^{2 \eta_2 y})} + 
\frac{D_0}{\sqrt{(1 - a^2 e^{2 \eta_1 y})(1 - a'^2 e^{2 \eta_2 y})}},\\
{\tilde{A_t}}^3_4 =C_0 \sqrt{(1 - a^2 e^{2 \eta_1 y})(1 - a'^2 e^{2 \eta_2 y})} - 
\frac{D_0}{\sqrt{(1 - a^2 e^{2 \eta_1 y})(1 - a'^2 e^{2 \eta_2 y})}},
\end{gathered}
\end{equation} 
where $C_0$ and $D_0$ are constants. 
In accordance with \eq{proj1}
the solution thus describes an M2-brane ending on  
two M5-branes with world-volumes spanning  
directions $1,2,3,4$ and $5,6,7,8$, respectively and sharing the 
directions $x$ and $t$ with the M2-brane \cite{jeon}. 
Thus we obtained a quarter-BPS configuration of the mass deformed BLG theory
given by a bound state of M2-M5-M5-branes.
\section{BPS configuration with $SO(2)\times SU(2)\times SO(4)$ symmetry}
\label{so4:sol}
In this section we consider $N=4$
BPS configurations in the modified BLG theory
with world-volume symmetry $SO(2)$. We find that upon choosing a certain
ansatz the  equations for the scalars reduce to the scalar equations of 
a self-dual $U(1)^2$ Chern-Simons theory with  two complex scalars.
Topological configurations in the latter case 
have been investigated in earlier
\cite{lpy,ccl,ziar}. However, no analytic solution to the equations 
appears to be known. 
We shall consider special cases in which we obtain certain solutions. 

As before,  to write down the $SO(2)$-invariant BPS equations we project the
supersymmetry variation of the fermion with the 
$SO(2)$ invariant BPS projector.
In terms of the $32 \times 32$ gamma matrices the projector is
\begin{equation}
\label{bpspro}
\Omega= \frac{1}{8} (1 + \alpha_1 \g^{x y}\G^{1 2})(1 + 
\alpha_2 \g^{x y}\G^{1 2})(1 + \alpha_3 \g^{x y}\G^{1 2})\ \mathcal{P}
\end{equation}
where $\alpha_1$, $\alpha_2$ and $\alpha_3$ are sign factors $\pm 1$ 
and $\mathcal{P}$ denotes the chiral projection 
matrix as before \eq{chiproject}. 
We shall consider the situation in which the 
$R$-symmetry is broken to $SU(2) \times SO(4)$. 
The projector \eqref{bpspro} assumes the form
\begin{equation}
\label{supro}
\Omega= 
\frac{1}{4}\big(1 + \g^{x y}(\G^{1 2} + \G^{3 4}) - \G^{1 2 3 4}\big)\ 
\mathcal{P}. 
\end{equation}
corresponding to the choice of the $\alpha$'s as 
$\{+,+,+\}, \{+,+,-\}$.
The projector corresponds to 
\begin{equation}
\label{proj2}
\g^{xy}\Gamma^{12}\theta = \g^{xy}\Gamma^{34}\theta =\theta, 
\end{equation} 
or, equivalently, $\Gamma^{1234}\theta = -\theta$. 
For simplicity, we  set the four scalars 
$X^5$, $X^6$, $X^7$, $X^8$ to zero and write the BPS equations for the
non-zero fields only. 
This reduction in the flavour degrees of freedom 
breaks the $R$-symmetry further to $SU(2)$.
The BPS equations in terms of the non-vanishing scalars, $X^I$, $I=1,2,3,4$,
are
\begin{equation} 
\begin{gathered}
\label{SOSU2}
D_x X^1 +  D_y X^2 = 0,\quad
D_x X^2 - D_y X^1 = 0,\\  
D_x X^3 +  D_y X^4 = 0,\quad
D_x X^4 - D_y X^3 = 0,
\end{gathered}
\end{equation} 
along with 
\begin{equation} 
\begin{gathered}
\label{SOSU21}
D_t X^1 + \ternary{X^1,X^3,X^4} + \eta_1 X^{2} = 0,\quad
D_t X^2 + \ternary{X^2,X^3,X^4} - \eta_1 X^{1} = 0,\\
D_t X^3 + \ternary{X^3,X^1,X^2} + \eta_1 X^{4} = 0,\quad
D_t X^4 + \ternary{X^4,X^1,X^2} - \eta_1 X^{3} = 0.
\end{gathered}
\end{equation} 
Using \eq{delY} these can be written in terms of the gauge-invariant
variables $Y$. From \eq{SOSU21}  we obtain
\begin{equation} 
\begin{gathered}
\label{Ytso21}
\pa_t Y^{11} + 2\eta_1 Y^{12} =0, \quad
\pa_t Y^{22} - 2\eta_1 Y^{12} =0, \\
\pa_t Y^{33} + 2\eta_1 Y^{34} =0, \quad
\pa_t Y^{44} - 2\eta_1 Y^{34} =0 
\end{gathered}
\end{equation} 
and
\begin{equation}
\label{Ytso22}
\begin{gathered}
\pa_tY^{12}+\eta_1(Y^{22}-Y^{11})=0,\quad
\pa_tY^{13}+\eta_1(Y^{23}+Y^{14})=0,\\
\pa_tY^{14}+\eta_1(Y^{24}-Y^{13})=0,\quad
\pa_tY^{23}+\eta_1(Y^{24}-Y^{13})=0,\\
\pa_tY^{24}-\eta_1(Y^{14}+Y^{23})=0,\quad
\pa_tY^{34}+\eta_1(Y^{44}-Y^{33})=0.
\end{gathered} 
\end{equation}  
These gauge-invariant equations provide restrictions on the choice of ansatz
for the scalar fields. 
For stationary configurations the time derivatives are set to zero and these
equations yield relations between the $Y$'s. 
In particular, they imply that $Y^{12}=Y^{34}=0$, meaning, $X^1,X^2$ are
mutually orthogonal, as are $X^3,X^4$. They also require $Y^{11}=Y^{22}$ and
$Y^{33}=Y^{44}$.
An ansatz satisfying these
relations compatible with the remaining $SU(2)$ $R$-symmetry is
\begin{equation} 
\label{xansatz1}
X^1 = \begin{pmatrix}f_1\\f_2\\0\\0\end{pmatrix},\quad
X^2 = \begin{pmatrix}f_2\\-f_1 \\0\\0\end{pmatrix},\quad  
X^3 = \begin{pmatrix}0\\0\\g_1\\g_2\end{pmatrix},\quad 
X^4 = \begin{pmatrix}0\\0\\ g_2\\-g_1\end{pmatrix}.
\end{equation} 
For this choice
For this ansatz the gauge group breaks down to $U(1)^2$. 
We retain the ansatz (\ref{gfield}) for the gauge field $\tilde{A_{\mu}}$. 
Let us introduce complex combinations, 
\begin{equation} 
\begin{gathered}
\label{complexworld}
z = x + i y,\\
\phi = f_1 + i f_2, \quad
\chi = g_1 + i g_2,\\
\tilde{A_z}^a_b = \tilde{A_x}^a_b - i \tilde{A_ y}^a_b.
\end{gathered}
\end{equation} 
We also define 
$A_z = \tilde{A_z}^1_2$ and $B_z=\tilde{A_z}^3_4$ 
for the spatial components of the gauge
field and 
$A_t = \tilde{A_t}^1_2$ and $B_t=\tilde{A_t}^3_4$ for the temporal parts. 
The gauge-invariant variables are, then 
\begin{equation}
Y^{11}=Y^{22}=|\phi|^2,\quad Y^{33}=Y^{44}=|\chi|^2 ,\quad Y^{IJ}=0
\,\,\text{for}\,\, I\neq J. 
\end{equation} 
In terms of these complex quantities the BPS equations \eqref{SOSU2} are
written as
\begin{equation} 
\begin{split}
\label{compBPS1}
D_z \phi &= \partial_z \phi + i  A_z \phi = 0,\\
D_z \chi &= \partial_z \chi + i B_z \chi = 0,
\end{split}
\end{equation}  
which can be solved to express  the gauge fields to be expressed as 
in terms of the scalars as
\begin{equation} 
\begin{split}
\label{spacegauge}
A_z &= i \partial_z \ln \phi,\\
B_z &= i \partial_z \ln \chi. 
\end{split} 
\end{equation} 
Similarly, assuming stationarity the equations
\eqref{SOSU21} relate the temporal part of the gauge field 
to the complex scalars $\phi$ and $\chi$, namely,
\begin{equation} 
\begin{split}
\label{compSOSU2}
A_t &= \eta_1 - |\chi|^2,\\ 
B_t &= \eta_1-|\phi|^2. 
\end{split}
\end{equation} 
Using \eq{spacegauge} and \eq{compSOSU2} we can now rewrite the action
\eq{action1} and the Gauss constraint \eq{gaussfull} in terms of the complex
scalars $\phi$ and $\chi$. The Lagrangian reads
\begin{equation}
\begin{split}
\label{actionSO2}
\mathcal{L} = &- 2 \big(|\partial_z \phi|^2 -  |\partial_z \chi|^2 + 2 |\chi|^2
(\eta_1 - |\phi|^2)^2 + 2 |\phi|^2 (\eta_1 - |\chi|^2)^2 \big),   
\end{split}
\end{equation}
while the Gauss constraint leads to two differential equations for the
scalars, namely, 
\begin{equation} 
\begin{split}
\label{vortexdiff}
\nabla^2 \ln|\phi|^2 - 2 |\chi|^2(|\phi|^2-\eta_1) = 0,\\ 
\nabla^2 \ln|\chi|^2 - 2 |\phi|^2(|\chi|^2-\eta_1) = 0,
\end{split}
\end{equation}  
where $\nabla^2 = \pa_z\pa_{\bar{z}}$.
Defining rescaled fields 
\begin{equation}
\tilde{\phi}= \phi / \sqrt{\eta_1},
\quad 
\tilde{\chi}=
\chi / \sqrt{\eta_1},
\end{equation} 
these two equations take the form 
\begin{equation} 
\begin{split}
\label{vortexdiff2}
\nabla^2 \ln|\tilde{\phi}|^2 
+ \lambda (1-|\tilde{\phi}|^2)|\tilde{\chi}|^2  = 0,\\ 
\nabla^2 \ln|\tilde{\chi}|^2 
+ \lambda (1-|\tilde{\chi}|^2)|\tilde{\phi}|^2  = 0,
\end{split}
\end{equation}  
where $\lambda = 2\eta_1^2$.
These coupled elliptic partial differential equations have been studied in the
context of $U(1)^2$ 
self-dual Chern-Simons theory with two complex scalar fields 
\cite{lpy,ccl}.
In particular, existence 
of topological vortex solutions, characterized by 
the boundary conditions 
\[\tilde{|\phi|}\rightarrow 1,|\tilde{\chi}|\rightarrow 1,\]
as $|z|\rightarrow\infty$,
have been established \cite{lpy}. For a single vortex at the origin, the
solution is also proved to be unique \cite{ccl}. However, no explicit
analytic construction of vortex solutions seem to exist in literature. 
The conserved $R$-charges for the $SO(2) \times SU(2)$ BPS configuration are given by
\begin{equation} 
\begin{gathered}
\label{rcharge12}
R_{12} =  \int d^2x |\phi|^2(\eta_1 - |\chi|^2),\\
R_{34} =  \int d^2x |\chi|^2(\eta_1 - |\phi|^2),
\end{gathered}
\end{equation}  
while the total energy of the configuration is given by 
\begin{equation}
\begin{split}
\label{energy2}
\mathcal{E} = &\frac{1}{4} \int d^2x \big(|\partial_z \phi|^2 + |\partial_z \chi|^2 + 4(\eta_1 
- |\chi|^2)^2 |\phi|^2 + 4(\eta_1 - |\phi|^2)^2|\chi|^2\big).
\end{split}
\end{equation}
\subsection{A special case}
Let us consider the special case of the
BLG theory without the four-form field, corresponding to the Lagranigan
$\mathcal{L}_{\text{BLG}}$. 
Putting $\eta_1=0$ in \eq{vortexdiff} we obtain the Gauss constraint
equations for this case as
\begin{equation} 
\begin{split}
\label{vortexblg}
\nabla^2 \ln|\phi|^2 - 2 |\chi|^2|\phi|^2 = 0,\\ 
\nabla^2 \ln|\chi|^2 - 2 |\phi|^2|\chi|^2 = 0.
\end{split}
\end{equation}  
Subtracting these we obtain 
\begin{equation}
\label{vortsub}
\nabla^2\ln \frac{|\phi|}{|\chi|} = 0.
\end{equation} 
Adding the equations \eq{vortexblg}, on the other hand,
we obtain a Liouville-like equation for $\rho = |\phi\chi|^2$, namely 
\begin{equation}
\label{vortadd}
\nabla^2\ln\rho = 4\rho, 
\end{equation} 
which is solved by 
\begin{equation}
\label{vortrho}
\rho = \frac{1}{2}\left|\frac{d\xi/dz}{1-|\xi(z)|^2}\right|^2,
\end{equation} 
where $\xi(z)$ is an analytic function of $z$.
From \eq{vortsub} and \eq{vortrho} we conclude that both $|\phi|^2$ and
$|\chi|^2$ are proportional to 
$\left|\frac{d\xi/dz}{1-|\xi(z)|^2}\right|$, 
modulo analytic functions. Thus, $|\phi|^2$, $|\chi|^2$, hence $Y^{II}$,
$I=1,2,3,4$, are singular on the
curve $\xi(z)=1$. Given a $\xi$, this corresponds to two M2-brane spikes
extended along 1-2 and 3-4 directions 
corresponding to the two $U(1)$ factors of the gauge group on
the original M2-brane of the BLG theory lying on the $z$-plane
\cite{jeon}. 
\subsection{Numerical solution}
\begin{figure}[h]
\begin{pspicture}(-1.5,-.3)(5,2.2)
\psset{xunit=2mm,yunit=15mm}
\savedata{\mydata}[{{0,0},
{1.98019, 0.00241}, 
{2.97029, 0.01547}, 
{3.96039, 0.05839}, 
{4.95049, 0.16350}, 
{5.94059, 0.37096}, 
{6.93069, 0.66643}, 
{7.92079, 0.88337}, 
{8.91089, 0.96707}, 
{9.90099, 0.99129}, 
{10.8910, 0.99773},
{11.88118, 0.99940}, 
{12.87128, 0.99984}, 
{13.86138, 0.99995}, 
{14.85148, 0.99998}, 
{16.83168, 0.99999}, 
{18.81188, 0.99999}, 
{20.79207, 0.99999}} 
]
\psline{<->}(21,0)(0,0)(0,1.5)
\psline[linewidth=.5pt](-.2,1)(.2,1)  
\rput(-.7,1){$\scriptstyle 1$}
\rput(-.7,0){$\scriptstyle 0$}
\psline[linewidth=.5pt](10,.05)(10,-.05)   
\rput(10,-.15){$\scriptstyle 10$}
\rput(0,-.15){$\scriptstyle 0$}
\dataplot[linecolor=darkgray,plotstyle=curve,showpoints=false]{\mydata}
\rput(20,-.2){$r$}
\rput(-1.7,1.4){$\scriptstyle |\phi|$}
\end{pspicture}
\hskip 1cm
\begin{pspicture}(-1,0)(5,2)
\psset{xunit=2mm,yunit=2mm}
\savedata{\mydata}[
   {{0.49504,9.7819},
   {0.99009,3.91276},
   {1.48514,2.51533},
   {1.98019,1.86312},
   {2.47524,1.48151},
   {2.97029,1.22876},
   {3.46534,1.04478},
   {3.96039,0.89586},
   {4.45544,0.75728},
   {4.95049,0.61072},
   {5.44554,0.45259},
   {5.94059,0.30036},
   {6.43564,0.17866},
   {6.93069,0.09763},
   {7.42574,0.05056},
   {7.92079,0.02542},
   {8.41584,0.01259},
   {8.91089,0.00620},
   {9.40594,0.00304},
   {9.90099,0.001494},
   {10.3960,0.00073},
   {10.8910,0.00036},
   {11.3861,0.00017},
   {11.8811,0.00008},
   {12.3762,0.00004},
   {12.8712,0.00002},
   {13.3663,0.00001},
   {13.86138,0.000005},
   {14.35643,0.000002},
   {14.85148,0.000001},
   {15.34653,0.0000006},
   {15.84158,0.0000003},
   {16.33663,0.0000001},
   {16.83168,0.00000007},
   {17.32673,0.00000004},
   {17.82178,0.00000002},
   {18.31683,0},
   {18.81188,0}}
]
\psline{<->}(22,0)(0,0)(0,13)
\dataplot[linecolor=darkgray,plotstyle=curve,showpoints=false]{\mydata}
\rput(20,-1.2){$r$}
\rput(-1.5,12){$\scriptstyle A_r$}
\psline[linewidth=.5pt](-.2,10)(.2,10)
\rput(-1.3,10){$\scriptstyle 10$}
\psline[linewidth=.5pt](10,.3)(10,-.3)   
\rput(10,-1){$\scriptstyle 10$}
\rput(-1,0){$\scriptstyle 0$}
\rput(0,-1){$\scriptstyle 0$}
\end{pspicture}
\caption{Plot of $|\phi(r)|$ and  $A_\theta$ with $\eta_1=1$}
\label{fig1}
\end{figure}
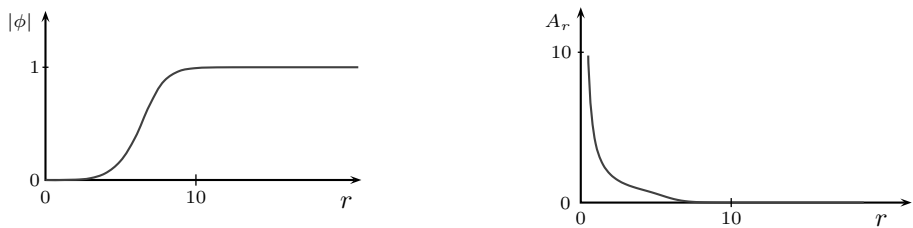

While general closed form solutions to \eq{vortexdiff2} are not known, we can
solve the equations numerically. Here we present a numerical solution for the
simplest case of a single vortex. Uniqueness of the solution for this case
has been established earlier \cite{ccl}.
Using the $SO(2)$ world-volume symmetry 
of the BPS configuration  let us now write 
$z = r e^{i\theta}$ 
and drop the $\theta$-dependence of all the functions. 
Writing $|\phi|^2 = e^{\rho(r)}$ and 
$|\chi|^2 = e^{\sigma(r)}$ the equations \eq{vortexdiff2} become
\begin{equation} 
\begin{gathered}
\label{diffr}
\rho''(r) + \frac{1}{r} {\rho'(r)} 
+ 2 e^{\sigma(r)}(\eta_1 - e^{\rho(r)}) = 0,\\ 
\sigma''(r) + \frac{1}{r}\sigma'(r) 
+ 2 e^{\rho(r)}(\eta_1 - e^{\sigma(r)}) = 0,
\end{gathered}
\end{equation}  
where a prime denotes a derivative with respect to $r$.
To solve these equations numerically to obtain vortex solutions
we have to impose two boundary conditions on each of 
$\rho$ and $\sigma$. 
The first set of asymptotic boundary conditions are chosen as 
$\rho(r) \rightarrow 0$ and $\sigma(r) \rightarrow 0$ 
as $r\rightarrow \infty$.
The second set of boundary conditions arise from the quantization of 
magnetic fluxes corresponding to the gauge 
fields $A_z$ and $B_z$, namely,
$\int dz \wedge d{\overline{z}} F_{z \overline{z}} = 
2 \pi N_1$ and 
$\int dz \wedge d{\overline{z}} F'_{z \overline{z}} = 2 \pi N_2$,
where $N_1$ and $N_2$ are integers representing
vorticity. In the limit $r \rightarrow \infty$
the flux quantization conditions written in terms of $\rho(r)$ and 
$\sigma(r)$ require
\begin{equation} 
\label{flux1}
|\rho'(r)|_{r \rightarrow \infty} = \frac{N_1}{r},\quad
|\sigma'(r)|_{r \rightarrow \infty} = \frac{N_2}{r}.
\end{equation} 
Solutions for $\rho$ and $\sigma$ can now be obtained numerically. A plot of
$|\phi|^2$ and the corresponding gauge field $A_\theta$(in polar cordinates) is shown in 
Figure~\ref{fig1} with unit vorticity for both vortices
and $\eta_1=1$. The plots of $|\chi|$ and $B_r$ are similar. 
\section{Summary}\label{concl}
To summarize, we have studied BPS configurations of the BLG theory
with and without the mass and four-form deformations. We considered three
cases of interest. In the first case the solution with world-volume symmetry 
$SO(1,1)$ preserving $N=(3,3)$
supersymmetry in the absence of any deformation has eight 
scalars which blow up at
a finite value of the world-volume coordinate $y$. We then considered a
quarter BPS configuration with $SO(4)\times SO(4)$ $R$-symmetry. 
In this case there are two types
of solutions. One of them is a pair of domain walls each extending along four
directions in agreement with the $R$-symmetry. The other solution features
the M2-brane merging into two M5-branes at a finite distance in $y$, the
latter intersecting along the $x$ direction. This has been interpreted as a
system of intersecting M2-M5-M5-branes \cite{berman}. Finally we considered
a configuration with $SO(2)$ symmetry in the world-volume and $SU(2)\times
SO(4)$ $R$-symmetry. We chose to turn off the four scalar corresponding to
the $SO(4)$. By choosing an appropriate ansatz for the scalars and the gauge
field, the system maps into the self-dual $U(1)^2$
Chern-Simons theory with two complex scalars. Existence of vortex solutions to
these equations has been established earlier. We presented a solution for the
special case with no deformation, giving rise to a Liouville-like equation.
We also presented a numerical solution for the single topological vortex,
which is known to be unique. 
In dealing with the system of BPS equations we found that expressing them in
terms of the gauge-invariant variables introduced earlier \cite{ckpr} appears
to be of immense help in the choice of ansatze for the solutions in all
cases. Other cases in  the classification of 
BPS configurations of the BLG theory may also be considered in a similar
fashion. However, the solutions for those are given either by constant
scalars or combinations of domain walls or the singular solutions of
section~\ref{so3so3sol}.
\section*{Acknowledgments}
We thank Pushan Majumdar, Krishnendu Sengupta, Subhashis Sinha for useful
discussions. 

\end{document}